\documentstyle[preprint,aps]{revtex}
\input{epsf}

\begin{document}
\draft
\title{Delocalization of tightly bound excitons in disordered systems}

\author
{Richard Berkovits}

\address{
The Minerva Center for the Physics of Mesoscopics, Fractals and Neural
Networks,\\ Department of Physics, Bar-Ilan University,
Ramat-Gan 52900, Israel}

\date{\today}
\maketitle

\begin{abstract}
The localization length of a low energy tightly bound 
electron-hole pair (excitons)
is calculated by exact diagonalization for small interacting disordered 
systems. The exciton localization length (which corresponds to the thermal
electronic conductance) is strongly enhanced by 
electron-electron interactions, while the localization length
(pertaining to the charge conductance) is only slightly enhanced.
This shows that the two particle delocalization mechanism widely discussed
for the electron pair case is more efficient close to the Fermi energy
for an electron-hole pair. The relevance to experiment is also discussed.
\end{abstract}


\narrowtext

\newpage

Interacting electrons
in random potentials are in the center of recent studies in
mesoscopic systems because of their 
relevance to many different phenomena such as 
the Coulomb gap\cite{es}, persistent currents \cite{gmw},
the two-dimensional (2D) MIT (metal-insulator transition) \cite{2dmit},
and delocalization of two particle states\cite{gmw,do,sh,imry}.

Although the subject of two particle state delocalization due to 
electron-electron (e-e) interaction has been the center
of many recent studies\cite{gmw}, most of those studies concentrate on
the situation of two particles in an empty band of a one-dimensional (1D)
system. This is an idealized 
situation which is not very relevant to experimental systems. A more
relevant situation, i.e. two excited electrons above the Fermi sea, was 
considered in Refs. \cite{imry,vow}. In Ref. \cite{vow} it was shown that two 
electrons close to the Fermi energy show no enhancement in 1D systems. 
Nevertheless, it has been emphasized that the situation may be more favorable
for higher dimensions\cite{imry}, for which, up to now, only the empty band 
case was studied\cite{jac}.
Recently it has been pointed out \cite{bs1} that when the effect of the 
Coulomb gap is taken into account the two electron 
delocalization scenario becomes less likely.

In this letter we consider a different type of two particle delocalization,
namely the delocalization of a tightly bound electron-hole pair (strongly 
bound excitons). 
This exciton is  analogous to the classical compact
electron-hole pair excitations discussed in Ref. \cite{es}, which can transfer
energy across the system but not charge. 
This is different from most previous discussion of two particle delocalization
which were concerned with the charge conductance
of the system. 
Thus, delocalization of the electron-hole
will manifest itself in the enhancement of the heat conductance of the
disordered system, an effect which can be measured in the laboratory. 
Indeed, the relevance of the two particle delocalization scenario to 
electron-hole pairs has been noted by Imry\cite{imry}. In Ref. \cite{bs1}
it was suggested 
that not only is the two particle localization length enhancement
relevant to an electron-hole pair, it is also much more effective than 
for the two electron case, because of the
Coulomb enhancement of the exciton density.
This scenario will be verified in this letter by explicit calculation
of the localization length.
Moreover, the existence of delocalized excitons may have interesting 
consequences for low temperature hopping conductance, which will be discussed
in more details after
presenting the evidence for the exciton localization length
enhancement.

Our study is based on the following interacting many-particle
tight-binding Hamiltonian:
\begin{eqnarray}
H= \sum_{k,j} \epsilon_{k,j} a_{k,j}^{\dag} a_{k,j} - V \sum_{k,j}
(a_{k,j+1}^{\dag} a_{k,j} + a_{k+1,j}^{\dag} a_{k,j}) + h.c
+ H_{int},
\label{hamil0}
\end{eqnarray}
where
$\epsilon_{k,j}$ is the energy of a site ($k,j$), chosen
randomly between $-W/2$ and $W/2$ with uniform probability, and $V$
is a constant hopping matrix element.
The interaction Hamiltonian is given by:
\begin{equation}
H_{int} = U
\sum_{k,j>l,p} {{a_{k,j}^{\dag} a_{k,j}
a_{l,p}^{\dag} a_{l,p}} \over
{|\vec r_{k,j} - \vec r_{l,p}|/b}}
\label{hamil2}
\end{equation}
where $U=e^2/b$ and
$b$ is the lattice unit.

We consider $L_x \times L_y$ samples of size
$3 \times 2$, $3 \times 3$, $3 \times 4$ and
$3 \times 5$
with $M=6,9,12,15$ sites and $N=2,3,4,5$
electrons correspondingly (i.e., $1/3$ filling).
The calculation of the electron transmission required also the
calculation of the $N+1$ electron Hamiltonian for each case. 
The $m \times m$ Hamiltonian (where $m=(^M_N)$)
matrix is numerically
diagonalized and all the eigenvectors $|\Psi_l^N\rangle$ and eigenvalues
$E_l^N$ are obtained.
The strength $U$ of the interaction is varied between $0-30V$
and the disorder strength is chosen to be $W=30V$.
The results are averaged over $6000$,~$1000$,~$200$,~$25$ 
different realizations
of disorder for the $M=6,9,12,15$ samples.

\begin{figure}
\vspace{-1.5cm}
\centerline{\epsfxsize = 3in \epsffile{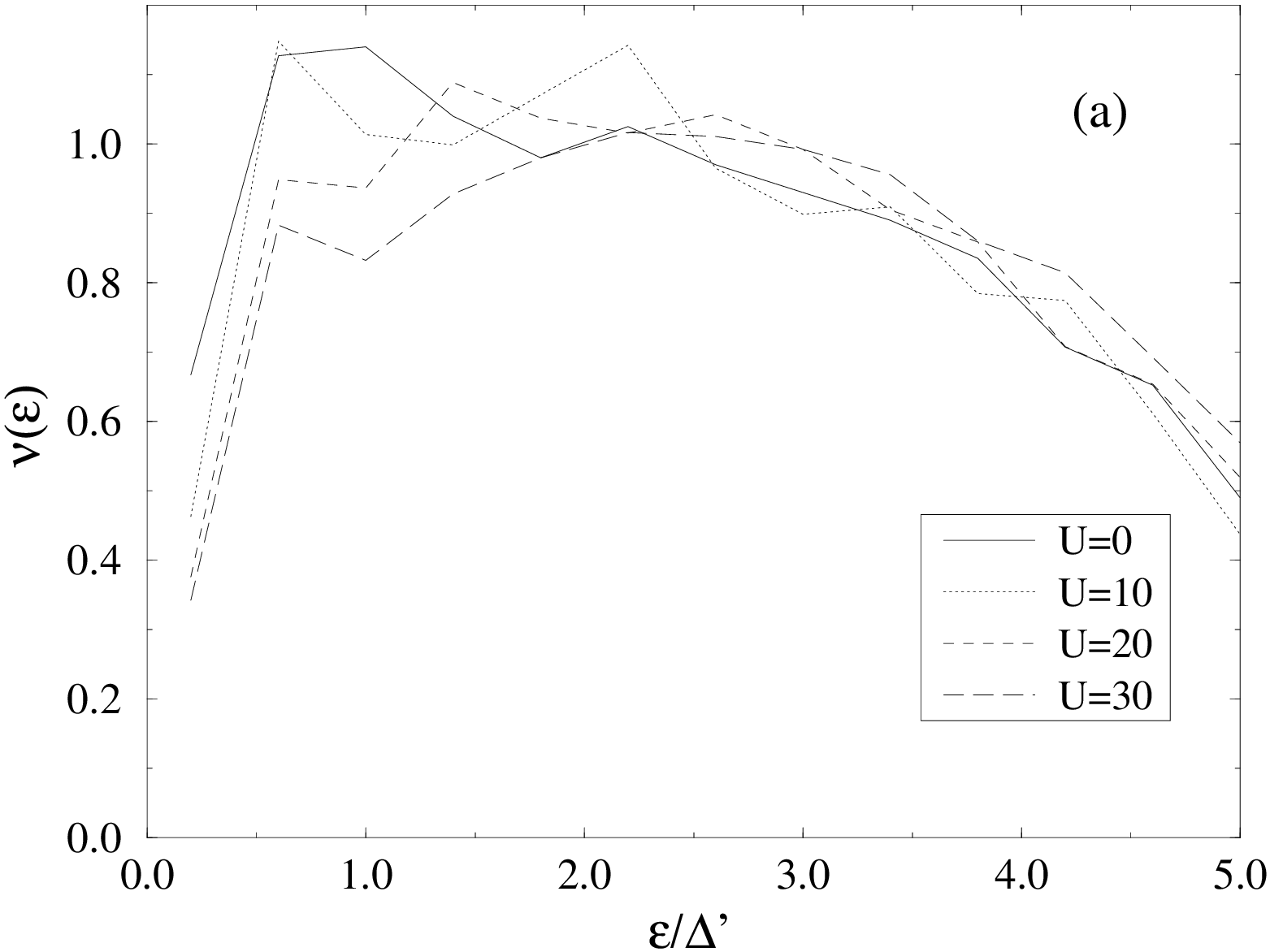}}
\vspace{-1cm}
\centerline{\epsfxsize = 3in \epsffile{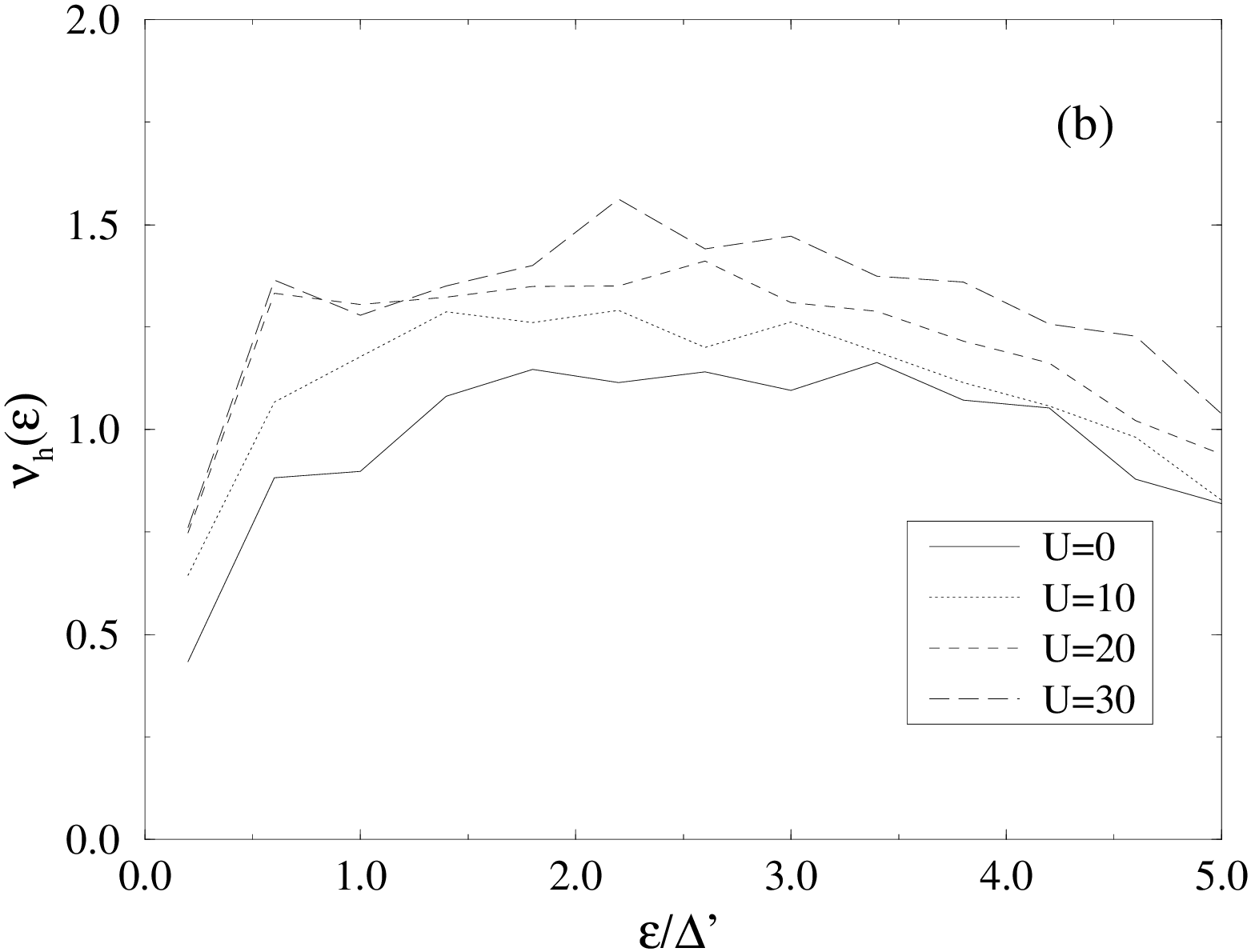}}
\caption{The average transport density of state (a) and average exciton
density of state (b) as function of the rescaled single electron level
spacing of a $3\times 3$ system
for different values of interaction strength.}
\label{fig1}
\end{figure}

The zero temperature
tunneling density of state (TDOS)
for the $N+1$ electron is defined in the usual way\cite{ep}:
\begin{eqnarray}
\nu(\varepsilon)=\sum_l | 
\langle\Psi_l^{N+1} | \sum_{k,j} a_{k,j}^{\dag}
|\Psi_0^N\rangle |^2 \delta(\varepsilon - (E_l^{N+1}-E_0^{N}))
\label{tdos}
\end{eqnarray}
which for the non-interacting case will reproduce the single particle
density of states above the Fermi energy. The averaged 
TDOS $\langle \nu(\varepsilon) \rangle = (0.4 \Delta')^{-1} 
\int_{\varepsilon-0.2\Delta'}^{\varepsilon+0.2\Delta'}
\langle \nu(\varepsilon') \rangle d \varepsilon'$ 
(here $\langle \ldots \rangle$
stands for an average over realizations of disorder)
for different values
of interaction is presented in Fig. \ref{fig1}a. Since the distance to the many
particle band edge $B$ increases
as function of the interaction strength $U$ we 
rescale the single particle level spacing $\Delta$ to offset this trend.
The rescaled ``single particle'' level spacing is defined as
$\Delta' = B/N(M-N)$, which is equal to $\Delta$ in the non-interacting 
case. For large values  $\Delta'$ there is little 
influence of the interaction strength on $\nu(\varepsilon)$, 
while for small values of 
$\Delta'$ there is a change. While for $U=0$ the TDOS
is rather monotonous except for a remanent of the level repulsion at 
$\Delta'=0$, there is a signature of the Coulomb gap \cite{es} 
at $U=30V$. Thus, the TDOS shows the expected behavior as function of the 
interaction strength.

We shall define a tightly bound exciton as the movement of an
electron from its ground state position
to a neighboring site, which may be expressed by the
application of a creation operator
$J_{k,j}^{\dag} = {J_{k,j}^{- x}}^{\dag}  + {J_{k,j}^{+ x}}^{\dag}
+  {J_{k,j}^{- y}}^{\dag}  + {J_{k,j}^{+ y}}^{\dag}$ (where 
${J_{k,j}^{\pm x}}^{\dag} = a_{k,j\pm 1}^{\dag} a_{k,j}$ and
${J_{k,j}^{\pm y}}^{\dag} = a_{k\pm 1,j}^{\dag} a_{k,j}$) 
on the ground state eigenvector.
The overlap of an excited state with a state obtained out
of the ground state by the creation of an exciton at site $(k,j)$
is then given by
$|\langle \Psi_l^{N+1} | J_{k,j}^{\dag} |\Psi_0^{N+1}\rangle |^2$.
One may also define the exciton density of state (EDOS)
\begin{eqnarray}
\nu_E(\varepsilon)=\sum_l | 
\langle \Psi_l^{N+1} 
| \sum_{k,j} J_{k,j}^{\dag} |
\Psi_0^{N+1}\rangle |^2 \delta(\varepsilon-(E_l^{N+1}-E_0^{N+1})))
\label{gdos}
\end{eqnarray}
in an analogous way to the TDOS. The EDOS then describes the weight of a
single tightly bound excitons at a given energy. The average EDOS
$\langle \nu_E(\varepsilon) \rangle$ (calculated in the same way as the
average TDOS)
for different values of $U$ is plotted in Fig. \ref{fig1}b.
It is clear that the interaction strength influences the EDOS
in a different way than the TDOS. There is no sign of of the Coulomb
gap, instead there is a monotonous enhancement of the EDOS, as
one may expect from the classical theory of tight bound excitons
which predicts a constant EDOS as function of $\varepsilon$ \cite{es}.

The localization properties of a system manifest themselves in the
transport behavior of the system. For a single particle system one
may connect the localization length $\xi$ to the transmission 
$t(\vec r,\vec r ~',\varepsilon) = \langle \vec r| ( \varepsilon -
H + i\eta)^{-1}|\vec r ~'\rangle$ via \cite{km}
$\xi(\varepsilon)^{-1}=\langle \ln |t(\vec r,\vec r ~',\varepsilon)| \rangle/
|\vec r - \vec r ~'|$. In Ref. \cite{mw} a many particle
Landauer formula connecting transmission and conductance for an interacting
system coupled to external leeds was developed. 
For the many particle case the transmission in the interacting region is
given by
\begin{eqnarray}
t(\vec r,\vec r ~',\varepsilon) = \sum_l
{{\langle \Psi_l^{N+1} | a_{\vec r}^{\dag} |\Psi_0^N\rangle
\langle \Psi_0^N | a_{\vec r ~'} |\Psi_l^{N+1}\rangle} \over{ 
\varepsilon - (E_l^{N+1}-E_0^{N})+i\eta}},
\label{t}
\end{eqnarray}
where $\eta \rightarrow 0$
In the numerical calculation 
we assumed $\vec r$ and $\vec r ~'$ to correspond to the opposite 
corners of the sample,
i.e., $\vec r= (1,1)$ and $\vec r ~' = (1,L_y)$.

The heat transport (namely the exciton transport) 
can be probed for example by attaching leads which excite
a tightly bound exciton at one point of the sample and absorbs it on another
point. In this case the heat transmission is formulated as
\begin{eqnarray}
t_h(\vec r,\vec r ~',\varepsilon)= \sum_l
{{\langle \Psi_l^{N+1} | J_{\vec r}^{\dag} |\Psi_0^{N+1}\rangle 
\langle \Psi_0^{N+1} | J_{\vec r ~'} |\Psi_l^{N+1}\rangle} \over{ 
\varepsilon-(E_l^{N+1}-E_0^{N+1})+i\eta}}.
\label{th}
\end{eqnarray}

\begin{figure}
\vspace{-1.5cm}
\centerline{\epsfxsize = 3in \epsffile{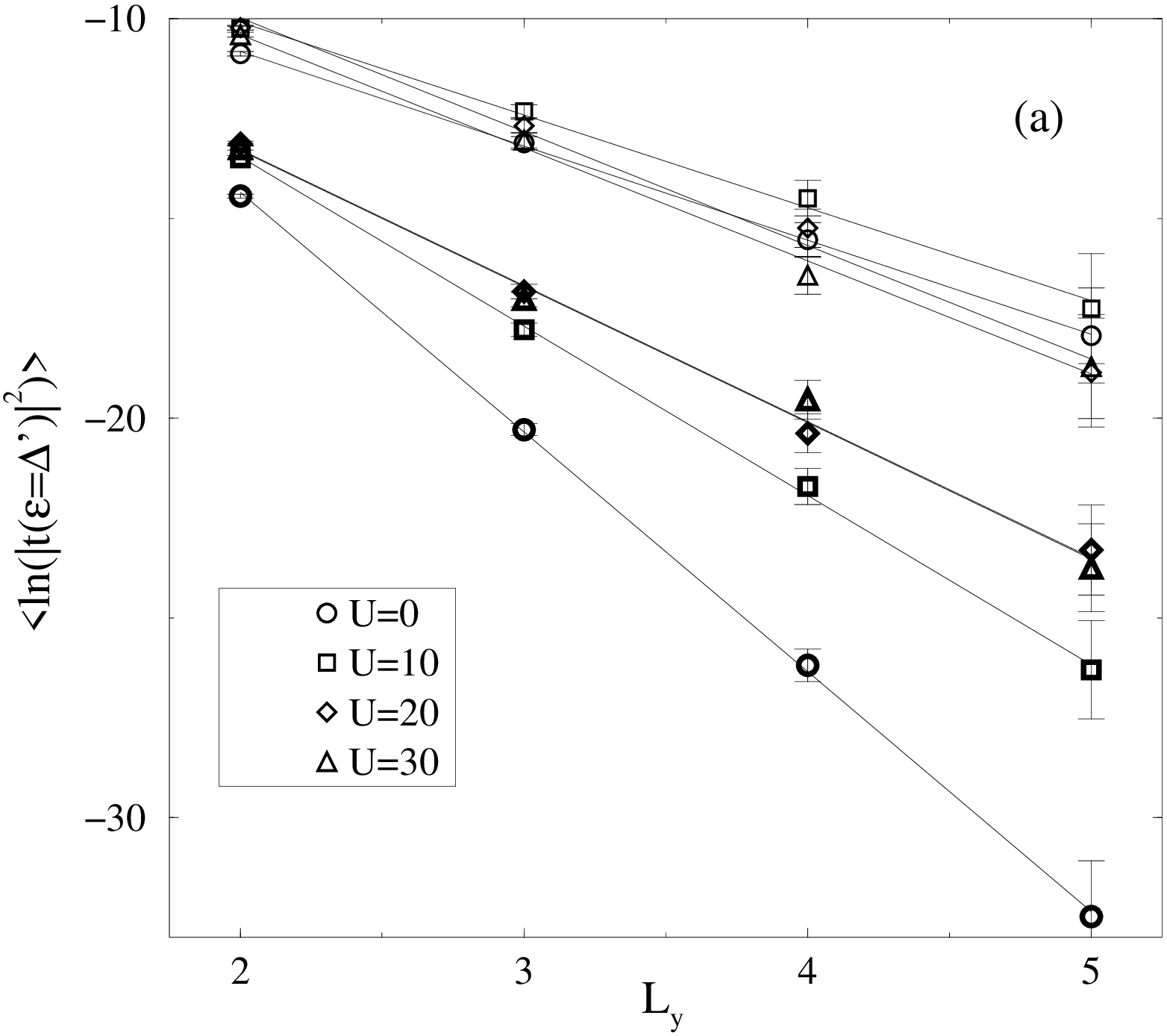}}
\vspace{-1cm}
\centerline{\epsfxsize = 3in \epsffile{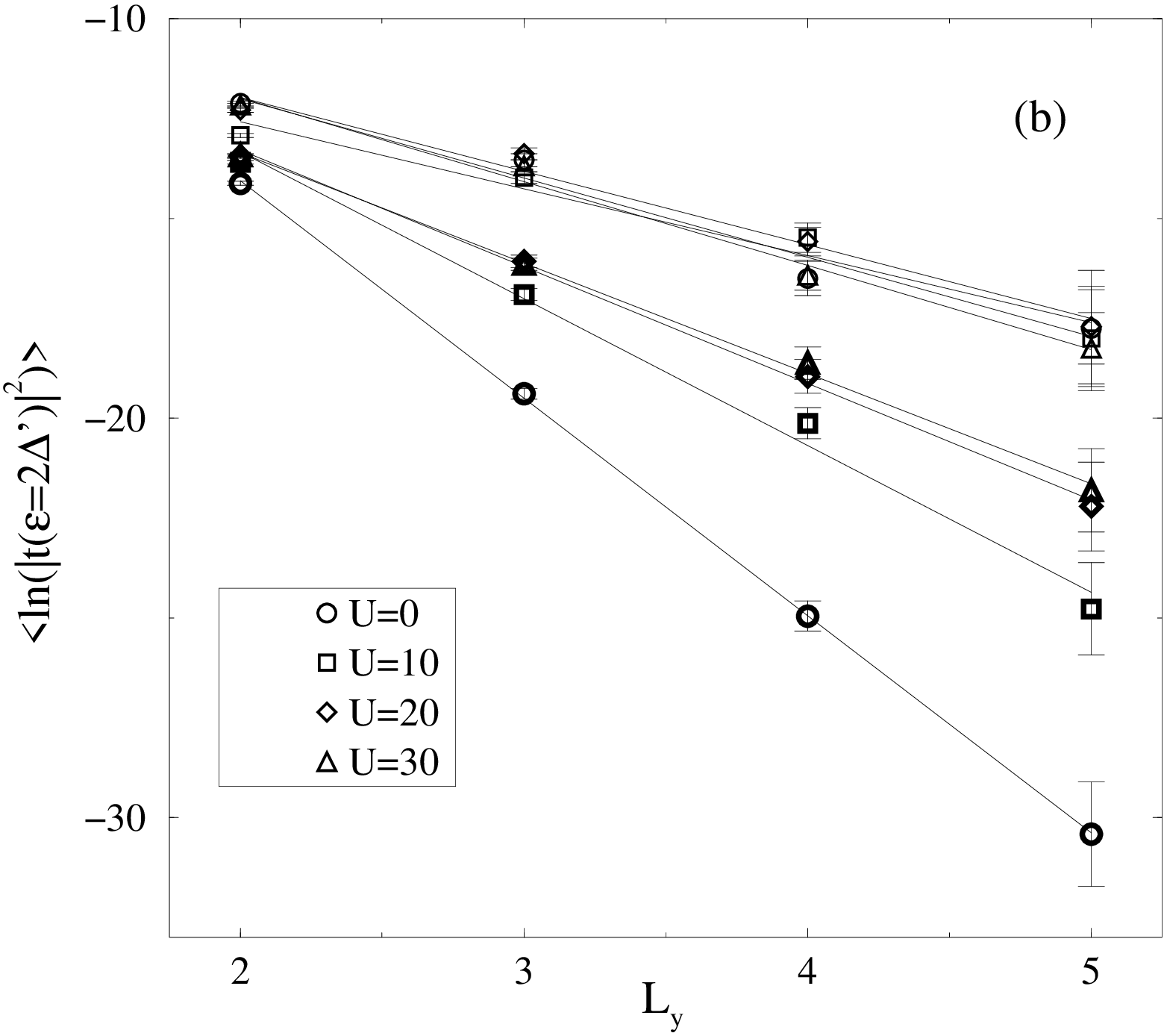}}
\caption{The average log of the charge (regular symbols) and heat
transmission (heavy symbols) as function of system length for
different values of interaction strength. (a) $\varepsilon=\Delta'$
 (b) $\varepsilon=2\Delta'$}
\label{fig2}
\end{figure}

The localization length of the system are extracted from the numerical data
in the following way: We calculate
the average log of the 
(heat) transmission between the two
corners of the sample for the low lying excitations
$\langle \ln | t_{(h)}(\varepsilon) | \rangle = \langle \ln ( (2 \Delta')^{-1}
| \int_{\varepsilon-\Delta'}^{\varepsilon+\Delta'} 
t_{(h)}(\vec r,\vec r ~',\varepsilon') d \varepsilon' |) \rangle $.
The results are presented in Figs. 
\ref{fig2}a for $\langle \ln | t_{(h)}(\varepsilon=\Delta') | \rangle$
and in \ref{fig2}b for
($\langle \ln |t_{(h)}(\varepsilon=2\Delta') | \rangle$).
From the slope of $\langle \ln t \rangle$ vs. $L_y$ one can deduce the
localization length according to 
$\xi \propto L_y / \langle \ln t \rangle$ \cite{km}. 
For the charge transmission we obtain the following localization
length $\xi= 0.85b,~0.86b,~0.70b,~0.71b$ 
at $U=0,~10V,~20V,~30V$ for $\varepsilon=\Delta'$
and  $\xi= 1.01b,~1.20b,~1.08b,~0.95b$ 
at $U=0,~10V,~20V,~30V$ for $\varepsilon=2\Delta'$.
Thus, the charge transmission is somewhat enhanced for weak interactions,
but is suppressed for stronger interactions.
This is in line with our expectations on the influence of the interaction 
on the charge conductance \cite{ba,ves}. On the other hand 
for the heat transmission the exciton localization length
$\xi_h= 0.33b,~0.47b,~0.59b,~0.59b$ corresponding to 
$U=0,~10V,~20V,~30V$ for $\varepsilon=\Delta'$
and $\xi_h= 0.37b,~0.55b,~0.69b,~0.72b$ corresponding to 
$U=0,~10V,~20V,~30V$ for $\varepsilon=2\Delta'$.  
Thus, the heat transmission and the exciton localization length 
are significantly enhanced by the interactions.

Thus, low energy excitons show a significant enhancement of their 
localization length, while electrons in that region show only a moderate
enhancement. This can be understood as the result of the fact that for
electrons interacting via the
Coulomb interaction the joint density of states of two electrons drastically
decreases at
small energies due to the Coulomb gap in the one-electron density of
states. This leads to a suppression of the enhancement of the localization
length which strongly depends on the two particle density of states.
However, for excitons, 
Coulomb effects increase the density of states making these arguments
more plausible. This can be clearly seen for the stronger interaction strength
($U=20V,30V$) for which $\xi_h$ are strongly enhanced, 
while $\xi$ is suppressed
compared to the non-interacting value. Also from a direct calculation of the
two electron transmission, the suppression of the two electron localization
length close to the Fermi energy is evident.

An enhancement of $\xi_h$ will of course manifest itself in electronic
thermal conduction measurements of disordered systems. For a case
in which $\xi_h>\xi$, which has not been achieved for the parameters we have 
chosen in our model, but can not be out ruled 
for larger systems or at a different 
parameter range,
interesting behavior may emerge. For example, it is quite possible
that for $\xi \gg b$ the exciton localization length may become infinite
so that the electronic conductivity is exponentially small while the
electronic thermal conductivity changes as a power of temperature
~\cite{bur,bs1}.
Another interesting effect in such a regime is the possibility that
at low
temperatures the delocalized excitons may assist electron hopping much more
effectively than phonons \cite{bs1,PS,APS}. 
This will result in a prefactor of
the variable range hopping proportional or even equal
to a universal value $e^2/h$.
Such prefactors were observed in the QHE regime (see references and analysis in
Refs. \cite{PS,APS}) and recently at zero magnetic field\cite{exp}.

In conclusion, we have shown that the two particle delocalization 
mechanism close to the Fermi energy is much more efficient for an 
electron-hole pair than for two electrons (or holes). This is a result of the
enhancement of the exciton density of states near the Fermi energy by Coulomb
interaction opposed to the suppression of the electronic density of states.
This results in
a large enhancement of the electronic thermal conductance 
while the conductance is only 
moderately enhanced.

I would like to thank B. I. Shklovskii for initiating this study and for many
important discussions and suggestions.
This research was supported by The Israel
Science Foundations Centers of
Excellence Program.

\end{document}